\documentclass[twocolumn, times, tighten]{aastex63}
\usepackage{graphicx,multirow,hyperref,url,color,xspace}

\newcommand{\msun}{{\rm M}_{\sun}}
\newcommand{\ledd}{L_{{\rm Edd}}}
\newcommand{\rg}{{R_{\rm{g}}}}
\newcommand{\xmm}{{{XMM-Newton}}\xspace}
\newcommand{\xte}{{{RXTE}}\xspace}

\begin{document}
\defcitealias{Garcia16}{G16}
\newcommand{\garcia}{{\citetalias{Garcia16}}\xspace}
\defcitealias{Jiang19a}{J19a}
\newcommand{\jiang}{{\citetalias{Jiang19a}}\xspace}

\title{Two Major Constraints on the Inner Radii of Accretion Disks}
\shorttitle{Inner Radii of Accretion Disks}

\author{Andrzej A. Zdziarski}
\affil{Nicolaus Copernicus Astronomical Center, Polish Academy of Sciences, Bartycka 18, PL-00-716 Warszawa, Poland; \href{mailto:aaz@camk.edu.pl}{aaz@camk.edu.pl, bdemarco@camk.edu.pl}}
\author{Barbara De Marco}
\affil{Nicolaus Copernicus Astronomical Center, Polish Academy of Sciences, Bartycka 18, PL-00-716 Warszawa, Poland; \href{mailto:aaz@camk.edu.pl}{aaz@camk.edu.pl, bdemarco@camk.edu.pl}}

\shortauthors{Zdziarski \& De Marco}

\begin{abstract}
The Stefan-Boltzmann law yields a fundamental constraint on the geometry of inner accretion disks in black hole X-ray binaries. It follows from considering the irradiating flux and the effective temperature of the inner parts of the disk, which implies that a strong quasi-thermal component with the average energy higher than that of a blackbody at the effective temperature has to be present whenever relativistic Fe K fluorescence and reflection features are observed. The apparent absence of such quasi-thermal component with the color temperature of $\sim$1\,keV in high-luminosity hard states is not compatible with a strongly irradiated disk extending close to the innermost stable circular orbit. Instead, the disk should be either truncated at a relatively large radius or irradiated by a corona at a large height, which would reduce the effective temperature and bring it to an agreement with the data. We also study constraints on disk/corona models following from comparing the disk densities fitted in literature using variable-density reflection codes with those calculated by us from the ionization parameter, the luminosity and the disk inner radius. We find that the fitted densities are much higher/lower in the hard/soft state of binaries, implying significant problems with the used assumptions and methods. 
\end{abstract}
\keywords{Accretion (14); Non-thermal radiation sources (119); X-ray binary stars (1811)}

\section{Introduction}
\label{intro}

The hard X-ray spectra of hard-state black hole (BH) binaries and of many Seyfert active galactic nuclei (AGNs) are well described by Comptonization in hot plasma. However, the location of this plasma, either on the BH rotation axis, above the accretion disk, or within its inner truncation radius, has been a matter of intense debate (e.g., \citealt{Kara19,Mahmoud19}). Even in cases in which there is a consensus on the presence of a truncated disk, the value of the inner truncation radius, $R_{\rm{in}}$, is not well determined. During quiescence of transient accreting BH binaries, $R_{\rm{in}}\gtrsim 10^4R_{\rm{g}}$ \citep{DHL01, Bernardini16}, where $R_{\rm{g}}=GM/c^2$ is the gravitational radius and $M$ is the BH mass. This implies that the disk is also highly truncated during the initial phases of the outburst. However, we still do not know when the disk reaches the innermost stable circular orbit (ISCO), whether already in the hard state or only during the transition to the soft state.

The currently prevailing view is, however, that above $\sim$1\% of the Eddington luminosity in the hard state the disk does extend very close to the ISCO and is strongly irradiated by either a compact corona or a compact source on the BH rotation axis (a lamppost) very close to the horizon (e.g., \citealt{Tomsick08, Reis08, Reis10, Furst15, Garcia15, Garcia18, Garcia19}). For example, \citet{Parker15}, based on the relativistic broadening of the Fe K complex, found that the disk in the hard state of Cyg X-1 extends to 1.1 of the ISCO radius and is irradiated by a lamppost located at a height $<$1.2 of the horizon radius of an almost maximally rotating BH. Those modeling also yielded measurements of the BH spin, and the presumed disk extending to the ISCO has then been used as a diagnostic putting constraints on alternative theories of gravity \citep{Xu18, Tripathi19, Zhang19a, Zhang19b, Wang20}.

Here, we derive two sensitive tests of the accretion geometry, and the truncation radius in particular. The first is based on the basic physical constraint that the pure blackbody emission is the strongest one achievable at a given source temperature (except for coherent radiation, which does not occur in hot conditions near accreting BHs). Since less effective thermalization results in the photon average energy higher than that of the blackbody, this sets a lower limit on the temperature of an irradiated disk, and, consequently, on $R_{\rm{in}}$. We then show that the observed absence of quasi-thermal spectral features with the effective temperature of $\sim$1 keV in luminous hard states implies that the inner disk is either absent or only weakly irradiated.

The second test follows from the X-ray reflection spectroscopy if both the ionization state of the reflector and its density can be determined. Since the ionization state depends primarily on the flux-to-density ratio, its fitted value together with the density yields the flux irradiating the reflector. If we also measure the observed flux and the distance to the source, this sets a constraint on the geometry of the primary X-ray source and the reflector.

\section{Thermalized Re-emission}
\label{thermal}

\subsection{Re-emitted spectra}
\label{spectra}

A major constraint on the source geometry in the hard state follows from considering the flux irradiating the cold medium in the system, $F_{\rm{irr}}$ (energy-integrated and per unit area). That flux is partly Compton-reflected, and partly absorbed and then re-emitted. If that re-emission would perfectly thermalize, it would appear as a blackbody at the effective temperature corresponding to the absorbed flux. However, the reprocessing does not lead to the full thermodynamic equilibrium, and the re-emitted spectrum contains lines and edges on top of a quasi-thermal continuum. Still, we have a strict lower limit on the shape of the re-emitted spectrum: its average energy has to be {\it above\/} the average energy of the blackbody at the effective temperature, $T_{\rm{eff}}$, given by the Stefan-Boltzmann law,
\begin{equation}
\sigma T_{\rm{eff}}^4=(1-a)F_{\rm{irr}}+F_{\rm{intr}},
\label{teff}
\end{equation}
where $\sigma$ is the Stefan-Boltzmann constant, $a$ is the albedo for backscattering and $F_{\rm{intr}}$ is the intrinsic flux generated by an internal dissipation. 

\begin{figure}
  \centering
  \includegraphics[width=7.5cm]{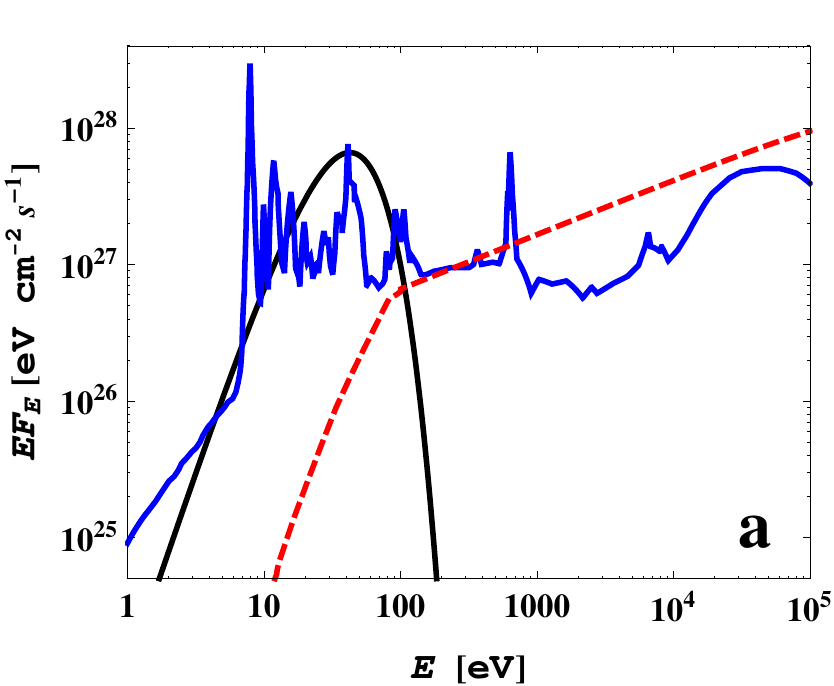}
  \includegraphics[width=7.5cm]{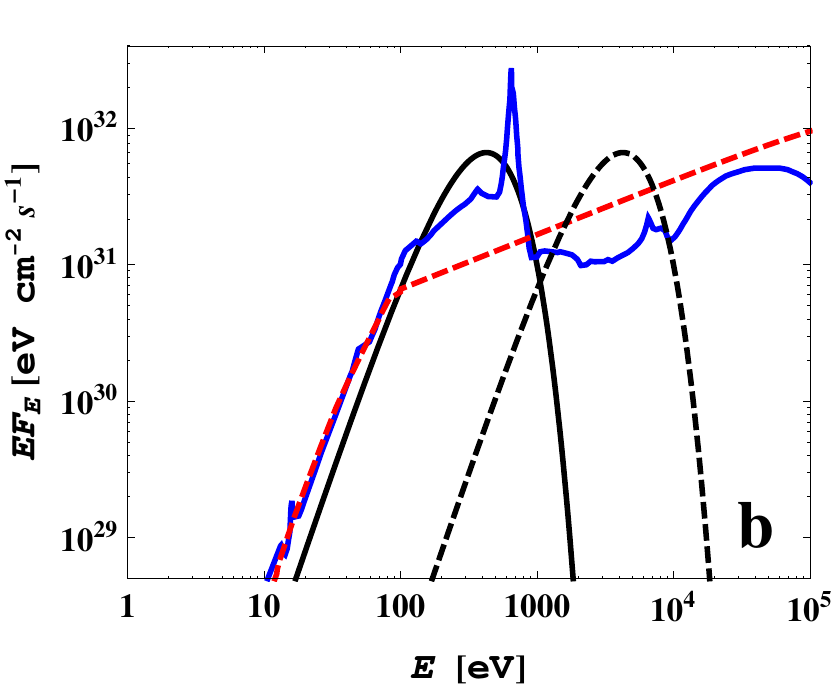}
  \includegraphics[width=7.5cm]{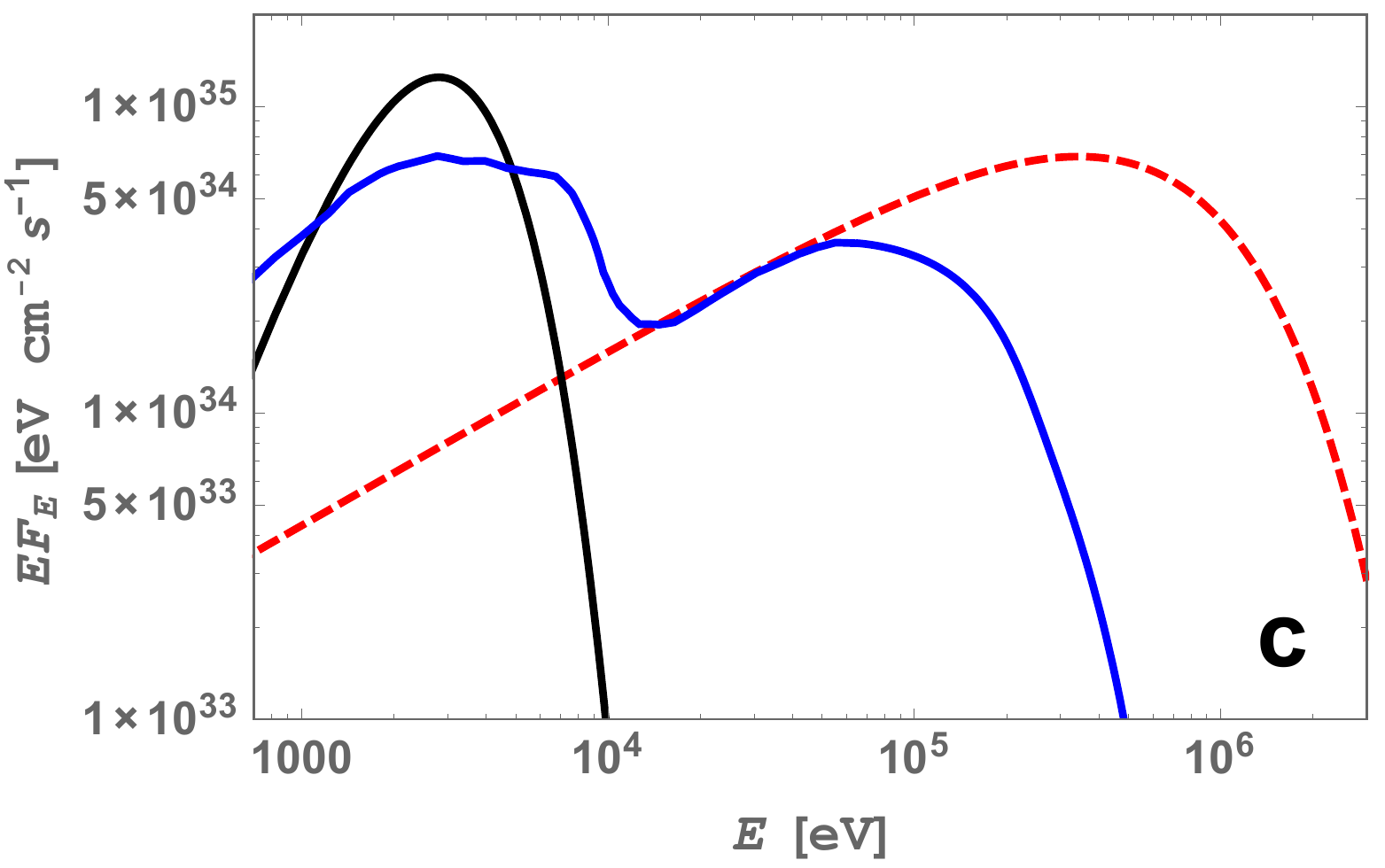}
  \caption{Angle-averaged re-emitted spectra (blue solid curves) from irradiation of strongly ionized constant-density slabs by incident e-folded power laws (red dashed curves) with a photon index, $\Gamma$, and an e-folding energy, $E_{\rm f}$. The black solid curves show the blackbody emission at the equivalent temperature, $T_{\rm eff}$. (a) The spectra from \garcia with the photon index $\Gamma=1.6$, the e-folding energy of $E_{\rm f}=1$\,MeV, the irradiating flux of $F_{\rm irr}\approx 4\times 10^{16}$\,erg\,cm$^{2}$\,s$^{-1}$, the electron density $n=10^{15}$\,cm$^{-3}$, and $kT_{\rm eff}\approx 10$\,eV. The prominent line at $\approx$650 eV is O{\sc viii} K$\alpha$. (b) As in (a) except for $F_{\rm irr}\approx 4\times 10^{20}$\,erg\,cm$^{2}$\,s$^{-1}$, $n=10^{19}$\,cm$^{-3}$, $kT_{\rm eff}\approx 110$\,eV. The black dashed curve shows the shape of the blackbody corresponding to the irradiating flux of $4\times 10^{24}$\,erg\,cm$^{2}$\,s$^{-1}$ ($kT_{\rm eff}\approx 1.1$\,keV). (c) The spectrum based on the spectral fit LF7 for GX 339--4 of \jiang. Here $\Gamma=1.427$, $E_{\rm f}=600$\,keV, $F_{\rm irr}\approx 4\times 10^{23}$\,erg\,cm$^{2}$\,s$^{-1}$, $n=10^{21}$\,cm$^{-3}$, and $kT_{\rm eff}\approx 720$\,eV.
}\label{spectra_G16}
\end{figure}

Figure \ref{spectra_G16} compares some published re-emitted spectra from irradiation with blackbody spectra at their corresponding $T_{\rm eff}$. They demonstrate that a blackbody can indeed roughly approximate the observed curvature of the reprocessed continuum, especially in cases with high irradiating fluxes. We show incident and re-emitted spectra for irradiation of strongly ionized slabs ($\xi\sim 10^3$\,erg\,cm\,s$^{-1}$) by an isotropic source above it, i.e., the reflection fraction (defined as the ratio of the irradiating flux to that emitted outside in a local frame) is ${\cal R}=1$. Since $F_{\rm{intr}}=0$ was assumed, the energy conservation requires then that both spectra contain the same energy-integrated flux. We use the re-emitted spectra from fig.\ 7 of \citet{Garcia16} (hereafter \garcia)\footnote{\citet{Garcia13} and \garcia report spectra in units of of erg\,cm$^{-2}$\,s$^{-1}$. However, this turned out to be a typo and the correct units are eV\,cm$^{-2}$\,s$^{-1}$. The e-folding energy in \garcia is given as $E_{\rm f}=300$\,keV, whereas it is $1$\,MeV; the ionization parameter is given as corresponding to the energy range of 0.0136--13.6\,keV whereas it is 0.1--$10^3$\,keV. Also, the re-emitted energy flux integrated over the 1--$10^6$\,eV range (with the $10^5$--$10^6$\,eV flux calculated by us using {\tt xillver}) for all five cases in fig.\ 7 of \garcia equals only $\approx$50\% (instead of $\approx$100\%) of the total incident flux. This is related to the non-relativistic treatment of Compton scattering (J. Garc{\'{\i}}a 2020, private communication). Given that energy conservation is crucial for our modeling, here $F_{\rm irr}$ is defined equal to the re-emitted flux.} calculated with their reprocessing code {\tt xillverD} and from fig.\ 6 of \citet{Jiang19a} (hereafter \jiang), their spectrum LF7 calculated with the reprocessing code {\tt reflionx} \citep{Ross99, Ross07} renormalized to ${\cal R}=1$. We compare the re-emitted spectra for the bolometric irradiating fluxes of $F_{\rm irr}\approx 4\times 10^{16}$, $4\times 10^{20}$ and $4\times 10^{23}$\,erg\,cm$^{-2}$\,s$^{-1}$ with those of a re-emitted blackbody component at $T_{\rm{eff}}$. The values of $F_{\rm irr}$ follow from the given values of the ionization parameter, $\xi$, defined as
\begin{equation}
\xi\equiv4{\pi}F'_{\rm{irr}}/n,
\label{xi}
\end{equation}
where $F'_{\rm{irr}}$ is the irradiating flux measured at the source in either the 0.01--100\,keV or 0.1--1000\,keV photon energy ranges and $n$ is either the H or electron density in the reflection codes {\tt{reflionx}}, and {\tt xillver}, {\tt{relxill}} and associated codes (\citealt{Garcia13}; \garcia; \citealt{Dauser16}), respectively. We note that the re-emitted spectra shown in Figure \ref{spectra_G16}, as well as those in \citet{Ross07} and \citet{Vincent16}, are for $kT_{\rm eff}\lesssim 1$\,keV. There remains an uncertainty about the form of the re-emitted spectra for hotter reprocessing media; they may become somewhat broader due to stronger efffect of Compton scattering in upper layers of the reflectors. However, as implied by our analysis below, values of $kT_{\rm eff}$ higher than $\sim 2$\,keV are not expected in the hard state of BH binaries.

We have approximately determined the values of the albedo based on the spectra (blue solid curves) as the fraction of the re-emitted flux in the range where the re-emitted spectrum is dominated by backscattering (rather than reprocessing). This happens above $E\approx 0.2$, 1.0 and 10\,keV in Figure \ref{spectra_G16}a, b, c, yielding $a\approx 0.73$, 0.66 and 0.34, respectively. While this is a crude method, the resulting blackbodies provide fair approximations to the overall shape of the quasi-thermal low-$E$ feature, especially in the high-flux cases b and c. On the other hand, a part of the emission at higher energies is from reprocessing as well, in particular O{\sc viii} K and Fe K complexes in Figures \ref{spectra_G16}a, b, and thus our albedo estimates are conservative. Furthermore, the treatment of \garcia (based on \citealt{GK10} and used in {\tt xillver} and {\tt xillverD}) does not correctly treat the Klein-Nishina effects (see footnote 1), which effects strongly reduce the albedo for hard spectra and the high-energy cutoffs at $\gtrsim$100\,keV, characteristic of the hard state. \citet{ZLG03} considered a spectrum from Comptonization with the photon index of $\Gamma=1.5$ and the electron temperature of $kT_{\rm e}=120$\,keV reflecting from a fully ionized medium, and found $a\approx 0.48$, much less than $a=1$ of the non-relativistic case, and not much larger than $a\approx 0.34$ corresponding to reflection of the same spectrum from a neutral medium. With the Klein-Nishina effects taken exactly into account, the values of the albedo for the cases of Figures \ref{spectra_G16}a, b would be reduced by $\sim$1/2 (because the present re-emitted spectra miss $\sim$1/2 of the incident flux, footnote 1). The results of \citet{ZLG03} also imply that even fully ionized disks in the hard state would show quasi-thermal features when irradiated.

Figures \ref{spectra_G16}b, c imply that the reprocessed spectral features in those high-flux cases have the average energy relatively close to the average energy of the blackbody at $T_{\rm{eff}}$, i.e., their ratio (the so-called color correction, $\kappa$; \citealt{ST95,Davis05}) is only slightly larger than unity. In particular, $\kappa\approx 1.3$ in the case c. Also, the seen broadening of that quasi-thermal feature as compared to the blackbody is similar to those of accretion disk atmosphere spectra from intrinsic dissipation \citep{Hubeny01}.

\subsection{Irradiating Fluxes}
\label{fluxes}

In order to use the above constraint, we need to estimate the irradiating flux. We use the observed bolometric flux, $F_{\rm{obs,0}}$, the distance to the source, $D$, and its characteristic size scale, $R$. Assuming isotropy of the irradiating (primary) source, its luminosity, $L_0$, equals $4\pi{D^2}F_{\rm{obs,0}}$. We initially consider a simple planar geometry with the primary source on each side of the disk emitting $L_0/2$ away from the disk and $L_0/2$ toward it (thus, the total luminosity is $L=2L_0$). The emission toward the disk is either Compton backscattered or absorbed and re-emitted, and each side of the disk emits in steady state $L_0/2$. The flux irradiating the disk can be written as $F_{\rm{irr}}=L_0/({\zeta}R^2)$, where $\zeta$ is a geometry-dependent factor. For example, for two isotropic point sources (lampposts) irradiating the disk at a distance $R$ at an angle $\theta$ between the ray and the disk normal, $\zeta=2\pi/\cos\theta$. For an isotropic corona above and below a ring between the inner radius $R$ and $R+\Delta R$, $\zeta\approx 4\pi\Delta{R}/R$. Then we consider a radially stratified isotropic and optically thin corona above and below the disk. We point out that unless the coronal scale height is large compared to the radius, we do not need to take it into account in the estimates since the corona emits a half of its power toward the disk and a half away, regardless of the scale height. The irradiating flux at the inner radius, $R_{\rm{in}}$, can be expressed as 
\begin{eqnarray}
&F_{\rm{irr}}(R_{\rm{in}})=\frac{4\pi{D^2}F_{\rm{obs,0}}}{\zeta R_{\rm{in}}^2}= \frac{4\pi \ell_0 m_{\rm p}c^5}{\zeta r_{\rm{in}}^2\sigma_{\rm T} G M} \label{fdisk}\\
&\approx 4.6\times 10^{24}\,{\rm erg\,cm^{-2}\,s^{-1}}\frac{\ell_0/0.2}{(\zeta/2\pi)(M/10\msun) (r_{\rm{in}}/2)^2}, \label{f_estimate}
\end{eqnarray}
where $\zeta=2\pi$ for $F_{\rm irr}(R) {\propto}R^{-3}$, $\ell_0\equiv L_0/L_{\rm{Edd}}$, $L_{\rm{Edd}}$ is the Eddington luminosity (for pure H), $r_{\rm in}\equiv R_{\rm{in}}/R_{\rm{g}}$, $m_{\rm{p}}$ is the proton mass, and $\sigma_{\rm{T}}$ is the Thomson cross section. Profiles ${\propto}R^{-3}$ approximately correspond to both dissipation of the gravitational energy in a disk and disk irradiation by a central source. This implies that the flux irradiating the inner radius of a disk extending close to the ISCO of a fast-rotating $10\msun$ BH in the bright hard state ($20\%\ledd$ or more) is $\sim 5\times 10^{24}$\,erg\,cm$^{2}$\,s$^{-1}$, i.e., $\sim\!10^8$ higher than values of $F_{\rm irr}$ used for the luminous hard state if the standard models {\tt relxill} and the public version of {\tt reflionx} are employed, see Figure \ref{spectra_G16}a. This flux is also factors of $\sim\!10^4$ and $\sim$10 higher than that shown in Figures \ref{spectra_G16}b, c, respectively (though for some geometries $F_{\rm{irr}}$ needs to be multiplied by the reflection fraction ${\cal R}$, see below). This implies the presence of a quasi-thermal feature with $kT_{\rm{eff}}\approx 1$\,keV, as shown in Figure \ref{spectra_G16}b, c. Indeed, at $F_{\rm{intr}}=0$ and $\zeta=2\pi$, we have the inner color temperature of $T_{\rm{in}}{\equiv}\kappa T_{\rm{eff}}(R_{\rm{in}})$ given by
\begin{equation}
k T_{\rm in}\approx 3.1 \frac{\kappa(1-a)^{1/4}\ell_0^{1/4}}{r_{\rm in}^{1/2}(M/10\msun)^{1/4}}\,{\rm keV}.
\label{teff_bhb}
\end{equation}

We stress that $T_{\rm eff}$ depends mostly on the irradiating flux, with only a secondary dependence on the density through the albedo, $a$. For $F_{\rm irr}(R){\propto}R^{-3}$, $T_{\rm eff}{\propto}R^{-3/4}$, which smooths the quasi-thermal feature, and makes it resemble a disk blackbody spectrum (which has the same temperature profile). Given that the observed hard-state inner disk blackbody temperatures are in the $\approx$0.2--0.5\,keV range (e.g., \citealt{Basak16,Wang-Ji18}), we need at least $r_{\rm{in}}\gtrsim 10$ at $a\sim\!0.5$ and $\ell_0=0.1$.

These estimates are modified by GR, e.g., radiation of a lamppost is focused toward both the disk and the horizon. The focusing toward the disk enhances the irradiating flux near the inner edge \citep{Martocchia96} with respect to that of Equations (\ref{fdisk}--\ref{f_estimate}), making our constraint stronger. If the lamppost is at a large height, $H\gg\rg$, in which case the characteristic distance from the primary source to the reflector is $\sim\!{H}$ rather than $\sim\!{R_{\rm{in}}}$, the relativistic broadening of the fluorescent features would be only modest even if the disk extends to the ISCO, similarly to the case of a truncated disk \citep{Fabian14}. The corona can also be outflowing \citep{Beloborodov99}. In that case, the irradiating flux of Equation (\ref{fdisk}) will be multiplied by the reflection fraction, ${\cal{R}}<1$, resulting from the Doppler de-boosting, but the rest of the estimates will remain unchanged. The dissipation profile, $\propto R^{-q}$, can have the index, $q$, different from 3. If $q\lesssim 2$, the disk can extend to the ISCO with a weak relativistic broadening, similarly to the case of a lamppost at $H\gg\rg$. However, rather large values of $q$ are fitted in literature, e.g., $q>7.2$ in \citet{Garcia18}. Furthermore, there is strong evidence from timing that the hard-state accretion disks do have intrinsic dissipation (e.g., \citealt{WU09,Uttley11}). Both $q>3$ and $F_{\rm intr}>0$ would further increase $T_{\rm eff}$. 

Finally, the geometry can be different from either pure disk corona or lamppost. In particular, if the disk (with a corona) is truncated at a $R_{\rm{in}}>R_{\rm{ISCO}}$ (the ISCO radius), it is very likely that the flow below $R_{\rm{in}}$ is dissipative and hot, e.g., \citet{YN14}. Then the irradiating flux will be lower than that estimated above, and the observed reflection fraction will follow from the fraction of the emission still taking place in the corona above the disk, the fraction of the hot-flow emission irradiating the disk, and the degree of scattering of the reflection in the corona. 

Both the reflection and quasi-thermal features would then be partly Compton upscattered by the corona. The latter would result in the total spectrum consisting of the quasi-thermal feature (resembling a disk blackbody) and its Comptonization, as well as reflection components, the same as in the currently standard model fitted to hard-state data. The only difference would be the $kT_{\rm eff}$ of that feature, $\sim$1\,keV if the disk extends close to the ISCO in the luminous hard state, higher than the observed range of $\sim$0.2--0.5\,keV.

\subsection{An Example of GX 339--4}
\label{GX}

We illustrate the above constraint in the case of a luminous occurrence of the hard state in the BH X-ray binary GX 339--4. We study its brightest hard-state observation by \xmm, on 2010 March 28. We use the EPIC-pn data in the timing mode, fit the energy range of 0.7--10 keV and exclude the 1.75--2.35 keV range (to discard the range affected by calibration uncertainties), as in \citet{Basak16}. The \xmm observation was accompanied by a contemporaneous observation  by the Rossi X-Ray Timing Explorer (RXTE), for which only the  Proportional Counter Array (PCA) data are available. We fit jointly both data sets in order to better constrain the underlying slope of the spectrum and to estimate the total flux. However, as shown in \citet{Basak16}, there are significant differences in the residuals in the joint fit of these data. This appears to be due to the PCA spectral calibration being significantly different from that of the EPIC-pn at $E\lesssim$10 keV, while the two are in good agreement at higher energies (see fig.\ 3 in \citealt{Kolehmainen14}). Therefore, in our joint fit, we use the PCA data only at $E>9$\,keV. On the other hand, these data become very noisy at $E>35$\,keV, which we discard. 

We fit the joint data by the model {\tt{tbabs(diskbb+ reflkerr)}}. The first component accounts for the interstellar medium (ISM) absorption \citep{WAMC00}. The relativistic reflection model {\tt{reflkerr}} \citep{Niedzwiecki19} assumes the thermal Comptonization model of \citet{PS96} as the incident spectrum, and we use the option {\tt{geom=0}}, which corresponds to the primary source being isotropic in the local frame. We assume the dimensionless BH spin of $a_*=0.998$ (at which $R_{\rm{ISCO}}\approx 1.237 R_{\rm{g}}$), and the emissivity $\propto{R^{-3}}$ (used here for relativistic broadening only). That model uses the static reflection model {\tt{xillver}} \citep{Garcia13}, which assumes $n=10^{15}$\,cm$^{-3}$. At the fitted ionization parameter of $\xi\!\sim\!10^3$\,erg\,cm\,s$^{-1}$ (Table \ref{fits}), the irradiating flux is very low (Eq.\ \ref{xi}), at which the bulk of the reprocessed emission is at $\lesssim$0.2\,keV, see Figure \ref{spectra_G16}a. The much higher $F_{\rm{irr}}$ of our fitted model implies the presence of a reprocessed soft component at higher energies. Reflection models suitable for high density ($>\!10^{19}$\,cm$^{-3}$) in the disk would be needed in order to self-consistently account for that. However, such models are currently not publicly available (\jiang). Therefore, we use the {\tt diskbb} disk blackbody model \citep{Mitsuda84} to approximate the reprocessed soft component, with the temperature of blackbody seed photons for Comptonization equal to that at the disk inner radius. The relative similarity of the disk blackbody emission to that from reprocessing in an accretion disk is pointed out in Section \ref{fluxes}. The models are used within the X-ray fitting package {\sc{xspec}} \citep{Arnaud96}.

\setlength{\tabcolsep}{4pt}
\begin{table*}[ht!]
\centering            
\caption{The results of the spectral fit with {\tt{tbabs(diskbb+reflkerr)}}.}
\label{fits}      
\vskip -0.5cm                               
\begin{tabular}{ccccccccccccc}
\hline 
%\hline
$N_{\rm H}$ & $y$ & $kT_{\rm{e}}$ & $r_{\rm{in}}$ & $Z_{\rm{Fe}}$ & $i$ & 
$\mathcal{R}$ & $\log_{10}\xi$ & $N$ & $kT_{\rm{in}}$ & $N_{\rm disk}$ & $A_{\rm{PCA}}$ & $\chi^2/\nu$ \\
$10^{21}{\rm cm}^{-2}$ &  &keV & & & $\degr$ & 
 & erg\,cm\,s$^{-1}$  &  & keV & $10^4$  & & \\
\hline
$7.4_{-0.1}^{+0.2}$ & $1.25^{+0.01}_{-0.01}$ & $42_{-3}^{+2}$ & $15.3_{-2.7}^{+3.4}$ & $4.9_{-0.2}^{+0.1}$ & $34_{-2}^{+1}$ & $0.28_{-0.02}^{+0.01}$ & $3.40_{-0.02}^{+0.02}$ & 0.74 & $0.207_{-0.004}^{+0.003}$ & $7.2_{-0.4}^{+0.3}$ & $1.31_{-0.01}^{+0.01}$ & 172/186\\
\hline                    
\end{tabular}\\ 
\tablecomments{       
$y\equiv4{\tau}kT_{\rm{e}}/m_{\rm{e}}c^2$, $Z_{\rm{Fe}}\leq5$ is imposed, $N$ is the flux density of {\tt{reflkerr}} at 1 keV, $A_{\rm{PCA}}$ is the relative PCA/EPIC-pn flux normalization. The uncertainties are for 90\% confidence, $\Delta\chi^2\approx2.71$.}
\end{table*}
            
\begin{figure}
  \centering
  \includegraphics[height=6.3cm,angle=-90]{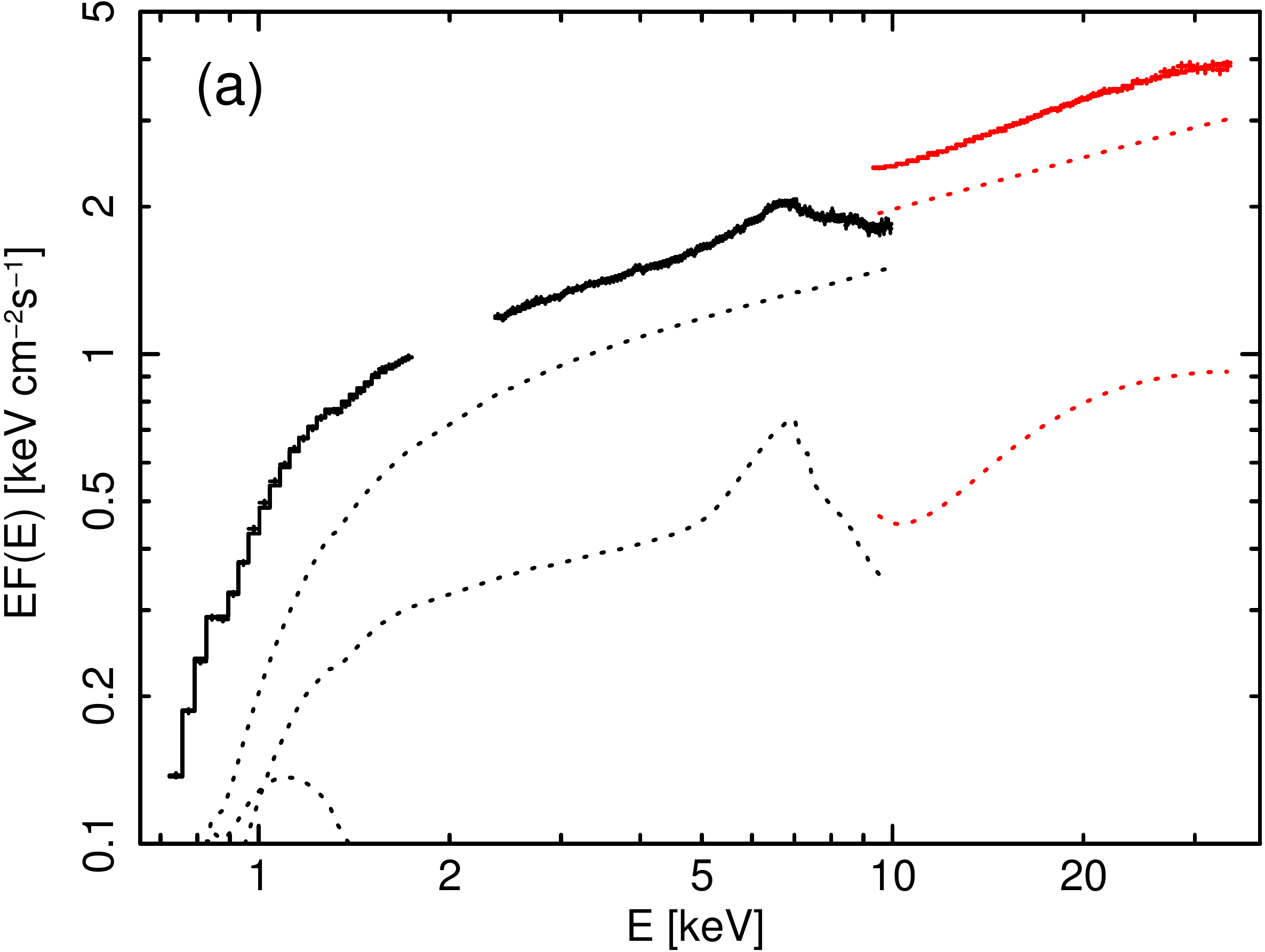}
  \includegraphics[height=6.3cm,angle=-90]{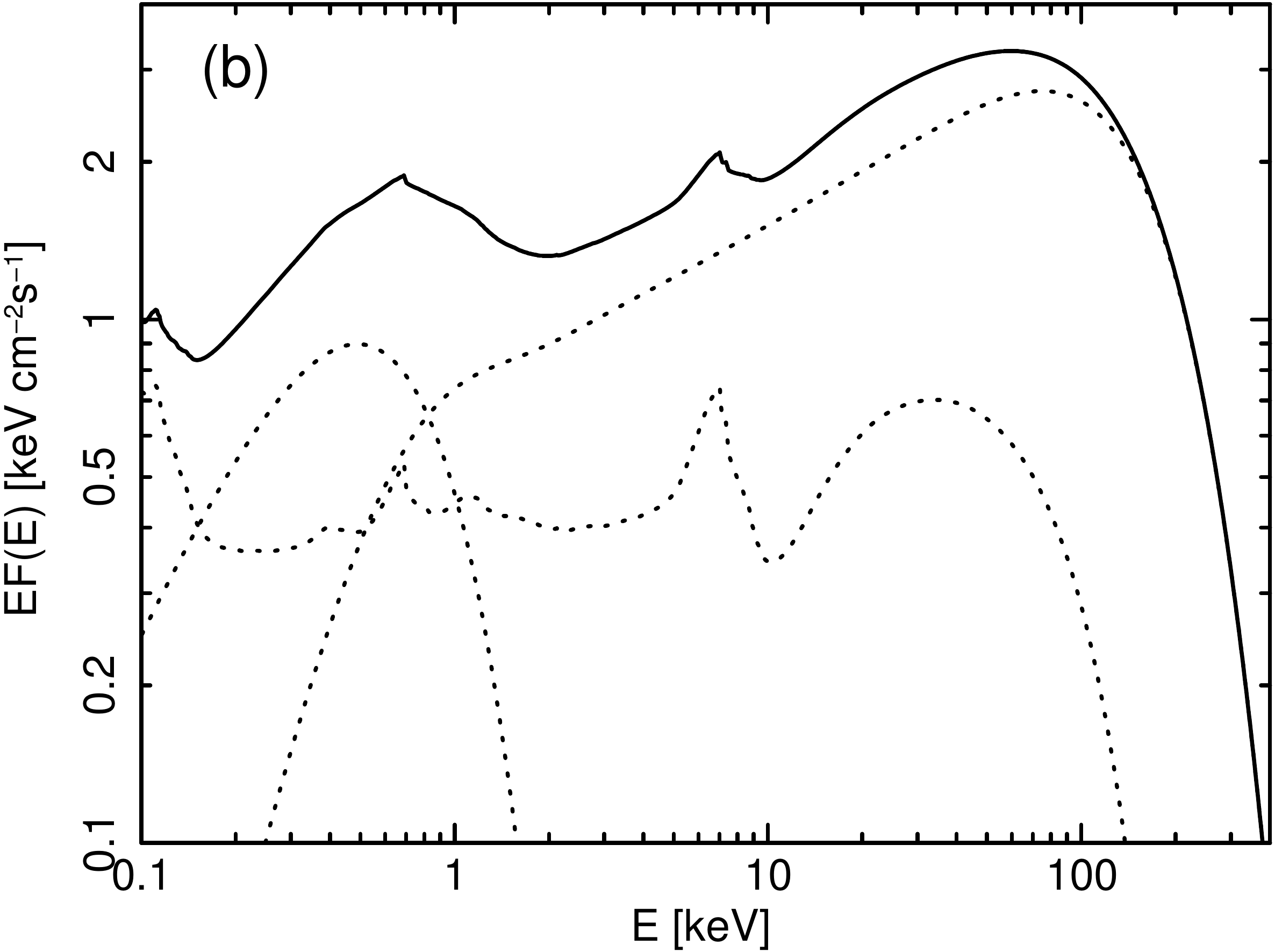}
  \caption{(a) Unfolded \xmm and \xte spectra. The dotted curves show the disk blackbody, the incident thermal Comptonization spectrum, and the reflected spectrum. (b) The unabsorbed model with the same components.
}\label{spectrum}
\end{figure}

Our fit results are given in Table \ref{fits}, the unfolded spectrum with the absorbed model and the unabsorbed model components are shown in Figure \ref{spectrum}a, b, respectively. We find a truncated disk, with $r_{\rm{in}}\approx 15$, and the reflection fraction of ${\cal{R}}\approx0.28$. The fitted high reflector Fe abundance, $Z_{\rm{Fe}}$, is probably an artefact of assuming a low disk density \citep{Tomsick18}. The fitted electron temperature of $kT_{\rm{e}}\approx 42$\,keV is similar to the values found in high-$L$ hard states of GX 339--4 by \citet{Wardzinski02}, whose fits included the  Compton Gamma Ray Observatory/Oriented Scintillation Spectrometer Experiment (CGRO/OSSE) data with significant source detection up to $\approx$500\,keV, thus well constraining the actual value of $kT_{\rm{e}}$.

We assume $M=8\msun$, $D=10$\,kpc (see \citealt{Heida17} and model D2 of \citealt{Zdziarski19}). The 0.1--$10^3$\,keV absorption-corrected flux is $\approx 2.4\times10^{-8}$\,erg\,cm$^{-2}$\,s$^{-1}$, which corresponds to the isotropic luminosity of $L\approx 2.9\times 10^{38} (D/10\,{\rm kpc})^2$\,erg\,s$^{-1}$, and 24\% of $L_{\rm Edd}$ at those $M$ and $D$. We estimate the flux from the \xmm data rather than from the PCA (which would yield the fluxes $\approx$1.3 times higher). The flux in the Comptonization component is $F_{\rm obs,0} \approx 1.6\times 10^{-8}$\,erg\,cm$^{-2}$\,s$^{-1}$. We estimate the albedo in the same way as for Figure \ref{spectra_G16}, obtaining $a\approx0.63$. This implies that the remaining flux has to be emitted as reprocessed emission at energies higher than those corresponding to $kT_{\rm{eff}}$, see Equations (\ref{teff}), (\ref{teff_bhb}).

\begin{figure}
  \centering
  \includegraphics[height=6.3cm,angle=-90]{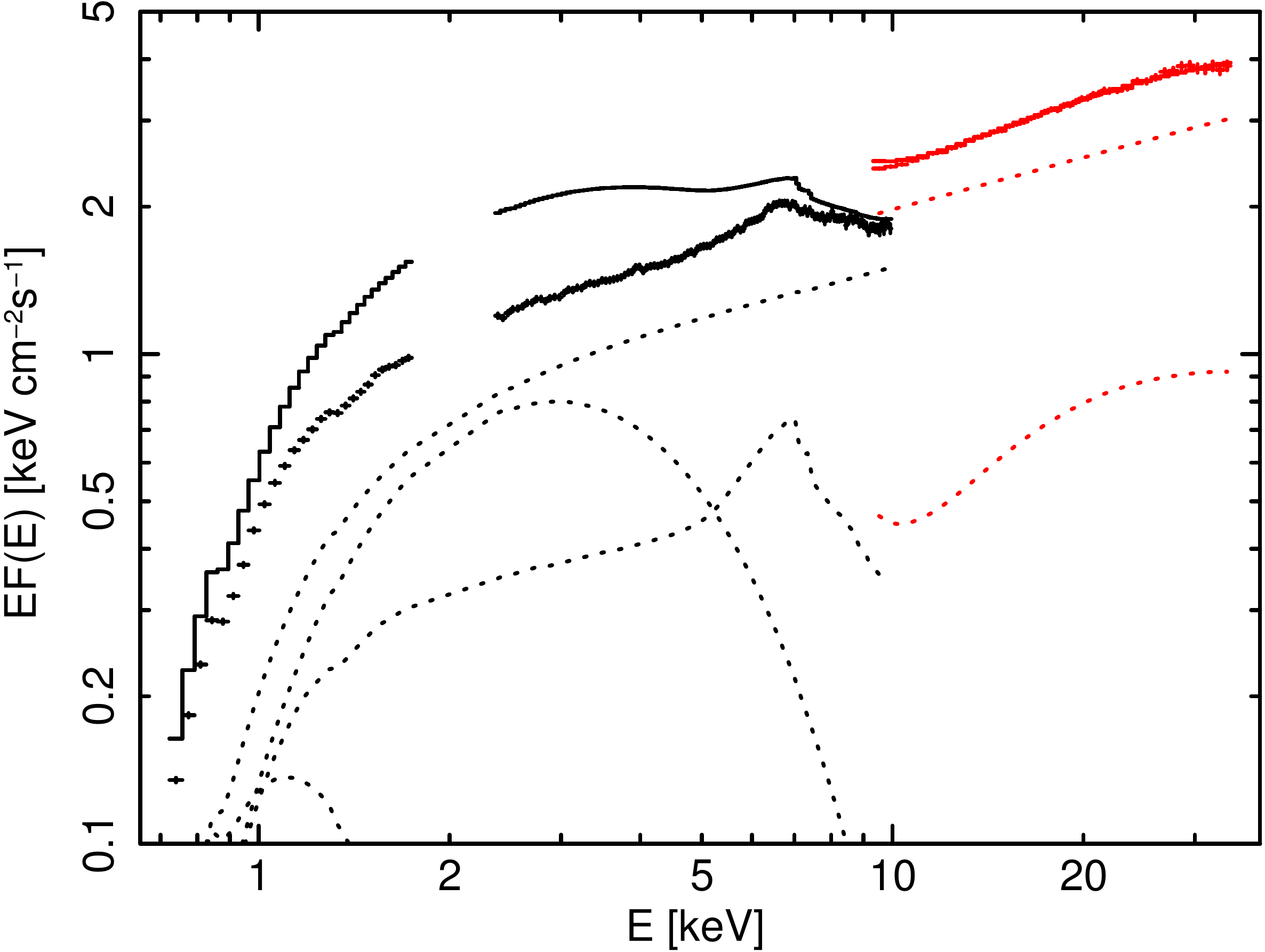}
  \caption{Unfolded \xmm and \xte spectra with the added disk blackbody component from irradiation with $kT_{\rm{in}}=1.12$\,keV and $F_{\rm{obs,disk}}=2.8\times{10^{-9}}$\,erg\,cm$^{-2}$\,s$^{-1}$ (at $i=34\degr$). The dotted curves show the two disk blackbodies, Comptonization and reflection. 
}\label{eeuf_bb}
\end{figure}

We then consider how to account for the obtained low fractional reflection, ${\cal{R}}\approx0.28$. One possibility is a reduction of the observed reflection due to scattering in the corona \citep{Steiner17}. We have tested this by Comptonizing a fraction, $f_{\rm{sc}}$, of the reflected spectrum using the convolution model {\tt ThComp} \citep{Z20}. The best fit was at a negligible scattering fraction, and $f_{\rm{sc}}\lesssim0.02$ at 90\% confidence. On the other hand, the fits to a spectrum from \xte PCA averaged over a few outbursts of GX 339--4 by \citet{Steiner17} using their code {\tt simplcut} allowed $f_{\rm{sc}}\lesssim0.5$ with ${\cal{R}}$ increasing then to $\sim$1. Thus, we consider this explanation still possible in principle. In that case, $F_{\rm{irr}}$ at the disk surface would be given by Equations (\ref{fdisk}--\ref{f_estimate}). 

Another possibility is that the primary source consists of two main parts; one forming a corona above the disk (and emitting $\sim$50\% toward it), and the other forming a hot flow inside the truncation radius, emitting mostly away from the disk, as for a partial overlap between hot and cold flows (cf.\ fig.\ 1f in \citealt{PVZ18}). The presence of the hot inner flow is supported by our fit with $R_{\rm{in}}{\gg}R_{\rm{ISCO}}$. Neglecting complications associated with details of this geometry, we assume that a fraction ${\cal{R}}$ of the primary luminosity, $L_0$, is emitted by the corona, whose reflection is then observed, and the remainder is emitted by the hot inner flow, without any associated reflection. We thus multiply $F_{\rm{irr}}$ of Equations (\ref{fdisk}--\ref{f_estimate}) by ${\cal{R}}$, which is a conservative choice, minimizing $T_{\rm{eff}}$. We further assume $F_{\rm{intr}}=0$, and obtain,
\begin{equation}
kT_{\rm{in}}\approx 1.12\,{\rm keV}\frac{\kappa}{1.3}\!\!\left(\frac{r_{\rm{in}}}{2}\right)^{-\frac{1}{2}}\!\! \left(\frac{D}{10{\rm{kpc}}}\right)^{\frac{1}{2}}\!\!\left(\frac{M}{8\msun} \right)^{-\frac{1}{2}}\!\!\!\!,
\label{teff_gx}
\end{equation}
where we scaled $T_{\rm{eff},in}$ to $r_{\rm{in}}=2$ similar to those obtained by \citet{Garcia15} for GX 339--4 and used $\kappa$ for the case of Figure \ref{spectra_G16}c. The observed disk blackbody flux in this component would then be
\begin{equation}
F_{\rm obs,disk}\approx 2\cos i (1-a){\cal{R}}F_{\rm{obs,0}},
\label{f_disk}
\end{equation}
where $i$ is the viewing angle (Table \ref{fits}), and the blackbody angular flux distribution is assumed. Then $F_{\rm obs,disk}\approx 3.3\cos i\times 10^{-9}$\,erg\,cm$^{-2}$\,s$^{-1}$, and $i\leq 60\degr$ is most likely \citep{Munoz13}, which constraint we impose. We thus added a disk blackbody component from the estimated irradiation in Figure \ref{eeuf_bb} in order to illustrate its appearance expected at a low $r_{\rm{in}}$. Since the observed spectrum shows almost no curvature above 2 keV, adding such a component is in a strong conflict with the data. We have still tried to fit it, by allowing $kT_{\rm in}$ and the flux of this component to be $\geq$1.12\,keV, $\geq\!3.3\cos i\times 10^{-9}$\,erg\,cm$^{-2}$\,s$^{-1}$, respectively, tying the temperature of the seed photons for Comptonization to $T_{\rm in}$ and allowing another disk blackbody at a lower temperature (which is strongly required by the data). The obtained fit was indeed very poor, with $\chi^2/\nu=5071/185$. Thus, the presence of such a high-$T_{\rm in}$ spectral component from irradiation appears incompatible with the data. This is also in agreement with previous fits of high-flux hard states of GX 339--4 \citep{Garcia15,Basak16,Dzielak19}, which show no trace of such a component. On the other hand, given that we have no access to a self-consistent spectral model for high illuminating fluxes (and high $n$), we do not attempt to set a quantitative limit on $r_{\rm{in}}$ for this observation. We can only state that $r_{\rm in}\gg 2$ is required for coronal models with the emissivity law $q\gtrsim 2$--3 or lampposts at heights less than a few tens of $R_{\rm g}$.

\section{The Disk Density}
\label{density}

A related major constraint follows if both the ionization degree, $\xi$, and the density, $n$, of the reflector are fitted to data (see also \citealt{Mastroserio19, Shreeram20}). Equation (\ref{xi}) implies $F_{\rm irr}'=\xi n/4\pi$, which irradiating flux can then be independently estimated from the observed direct flux, $F_{\rm{obs,0}}$, see Section \ref{fluxes}. In particular, $n$ and $\xi$ can be fitted using either {\tt reflionx} or {\tt relxillD}, yielding $n=n_{\rm{fit}}$ and its uncertainty, as done by \citet{Tomsick18} and \jiang; \citet{Jiang19b}. Here, we use their results to check the self-consistency of their models, with the goal of being able to constrain $r_{\rm{in}}$.

In those papers, the irradiating spectrum is an exponentially cut-off power law with a photon index, $\Gamma$, and the e-folding energy, $E_{\rm{f}}$. $F_{\rm{obs,0}}$ can be then rescaled to $F'_{\rm{obs,0}}$, the observed flux in the energy ranges used in either code using the fitted values of $\Gamma$ and $E_{\rm{cut}}$. Including ${\cal R}$ in Equations (\ref{fdisk}--\ref{f_estimate}) gives $n_{\rm{cal}}=16\pi^2{\cal{R}}F'_{\rm{obs,0}} D^2/(\zeta R^2\xi)$. We take $R=R_{\rm{in}}$ as fitted to the observed reflection spectra using radial emissivity profiles, giving the calculated $n$ as
\begin{equation}
n_{\rm{cal}}=\frac{16\pi^2{\cal{R}}F'_{\rm{obs,0}} D^2c^4}{\zeta r_{\rm{in}}^2 G^2M^2\xi}=\frac{16\pi^2{\cal{R}}\ell'_0 m_{\rm{p}}c^5}{\zeta r_{\rm{in}}^2 \sigma_{\rm T} G M \xi}, \label{n2}
\end{equation}
which, at the emissivity $\propto{R^{-3}}$ and $\zeta=2\pi$, yields
\begin{equation}
n_{\rm{cal}}\approx \frac{1.2\times 10^{24} {\cal{R}}\ell'_0 }{r_{\rm{in}}^2 (M/10\msun) (\xi/10^3{\rm erg\,cm\,s}^{-1})}\,{\rm cm}^{-3}. \label{n3}
\end{equation}
These densities are comparable to those of accretion disks with a part of the dissipation taking place in a corona (\citealt{SZ94}; \garcia).

\begin{figure}
  \centering
  \includegraphics[width=7cm]{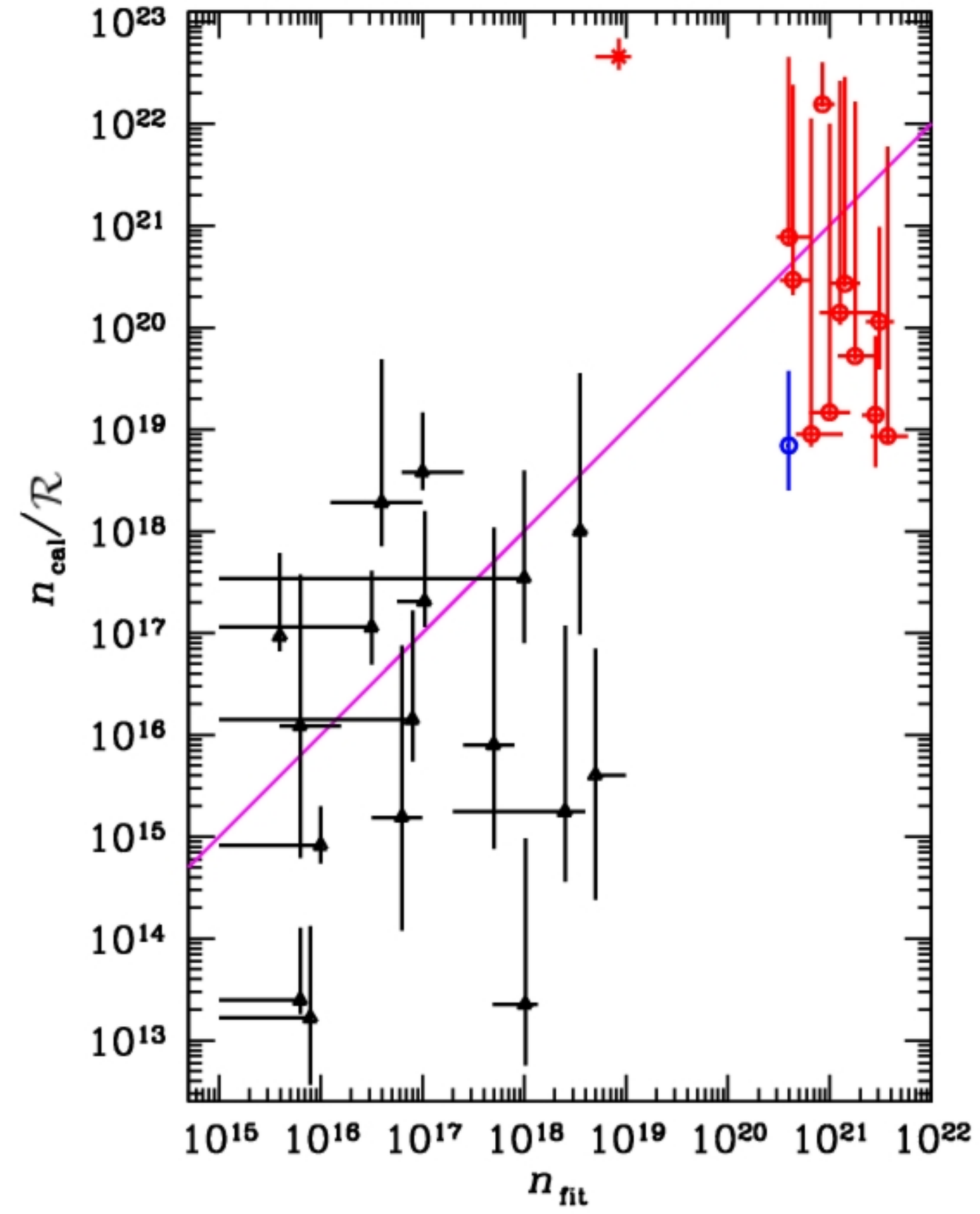}
  \caption{Comparison of the values of the density fitted by the reflection codes, $n_{\rm{fit}}$, and those based on the irradiating flux scaled to ${\cal{R}}=1$, $n_{\rm{cal}}/{\cal R}$, Equation (\ref{n2}). The magenta line shows $n_{\rm{cal}}/{\cal R} =n_{\rm{fit}}$. The red symbols, blue open circle and black triangles (all with error bars) show the results for GX 339--4, Cyg X-1 and Seyferts fitted by \jiang, \citet{Tomsick18} and \citet{Jiang19b}, respectively. The red open squares and the cross correspond to the hard and very high state, respectively, of GX 339--4.
}\label{ncal_nfit}
\end{figure}

The results for $n_{\rm cal}/{\cal R}$ from these calculations are shown in Figure \ref{ncal_nfit}. The magenta diagonal line corresponds to $n_{\rm{cal}}/{\cal R}=n_{\rm{fit}}$ at $\zeta=2\pi$, with ${\cal{R}}=1$ corresponding to the assumption of the purely coronal geometry. The uncertainties on $n_{\rm{cal}}$ are based on the uncertainties on the direct flux, $\Gamma$, $E_{\rm{cut}}$, $\xi$, $r_{\rm{in}}$, and $M$. Often, only upper or lower limits of some parameters are available. In those cases, we use the given limits in calculating the middle points of $n_{\rm{cal}}$, which results in some error ranges being strongly asymmetric. We note that these parameters have often large uncertainties, of the order of the given best-fit quantity or larger. Therefore, we cannot use the standard propagation of errors, since it assumes the uncertainty to be much lower than the best-fit value. Instead, we estimate the uncertainties by calculating the maximum and minimum of $n_{\rm{cal}}$ corresponding to the extreme values of the parameters that maximize and minimize, respectively, its value. This method is conservative, i.e., it gives errors larger than those resulting from random scatter of the parameters.

The red points refer to the 12 observations of GX 339--4 studied by \jiang using {\tt{reflionx}} convolved with {\tt{relconv}} \citep{Dauser16}, assuming $a_*=0.998$. We assume $M=8\msun$ and $D=10$\,kpc; uncertainties on them and on ${\cal{R}}$ and $\zeta$ can be accounted for by rescaling $n_{\rm{cal}}$ by a factor ${\cal{R}}D^2/(\zeta M^2)$. The points correspond to 11 observations in the low-flux hard state and one in the high-flux very high state (studied before by \citealt{Parker16}). We see that, at ${\cal{R}}=1$, most of the hard-state values of $n_{\rm{cal}}$ are compatible with being equal to $n_{\rm{fit}}$ within the error bars. The only clearly discrepant point with small error bars is the one characterized by the highest value of $n_{\rm{cal}}$ for the hard state (LF10 in \jiang), for which $R_{\rm{in}}<1.51 R_{\rm{ISCO}}$ was reported, while $R_{\rm{in}}\lesssim(5$--$30)R_{\rm{ISCO}}$ were found for the remaining 10 hard-state observations. In most cases, they reported only upper limits on $r_{\rm in}$ consistent with substantially truncated disks. 

However, these results are affected by the actual values of ${\cal{R}}$. While their best-fit values are not reported in \jiang, they can be approximately inferred from their fig.\ 6 as the ratio of the relativistic reflection component to the power-law one in the range of 10--30 keV for the angle-averaged emission. The resulting values are ${\cal{R}}\approx 0.015$--0.04 in all hard-state cases except observation LF1, where we infer ${\cal{R}}\approx 0.1$, i.e., are all $\ll$1. This explains the absence of a strong low-energy feature in the total model spectra, in spite of their dominance in the re-emitted spectra seen in Figure \ref{spectra_G16}c (based on the model for observation LF7, where we inferred ${\cal{R}}\approx 0.018$). ${\cal R}\ll 1$ cannot be explained by scattering in the corona because then the power-law component would be from Comptonization of the quasi-thermal feature, while the power law lies much above that feature for most of the spectra shown in fig.\ 6 of \jiang. ${\cal R}\ll 1$ cannot also be explained in the lamppost geometry, where gravitationally focused irradiation would result in ${\cal R}\gtrsim 1$, similar to that shown in Figure \ref{spectra_G16}c. Then, the model that can better explain ${\cal{R}}\ll 1$ is a truncated disk plus a hot inner flow, only weakly irradiating the disk (as discussed in Section \ref{GX}). Given ${\cal R}\ll 1$, the values of $n_{\rm{cal}}$ for most of the points plotted in Figure \ref{ncal_nfit} for the hard state of GX 339--4 are below $n_{\rm fit}$. We note that the models in \jiang also contain a luminous distant (not relativistically broadened) reflection component, which is in all of the cases either similarly strong or stronger than the relativistic component (accounting for most of the reflection humps around 30\,keV). Both the very low values of ${\cal R}$ and the dominance of remote reflection are in principle possible but differ from all previous fits to the hard state of GX 339--4 (e.g., \citealt{Garcia15, Basak16, Wang-Ji18, Dzielak19}). 

On the other hand, the very high state observation has small error bars and $n_{\rm{cal}}/{\cal{R}}\approx 5400 n_{\rm{fit}}$ at $R_{\rm{in}}= R_{\rm{ISCO}}$. This discrepancy could be removed for ${\cal{R}}^{-1/2} (\zeta/2\pi)^{1/2} (R_{\rm{in}}/R_{\rm{ISCO}}) (M/8\msun) (10\,{\rm{kpc}}/D) \approx 70$, which appears unlikely. Here, no distant reflection was included and we found ${\cal{R}}\sim 1$. This discrepancy might be a consequence of the adopted assumption of a passive accretion disk. Such an assumption may be approximately correct for the hard state, where most of the luminosity is emitted by the corona, but it is certainly not proper for the soft state, where the dominant disk intrinsic dissipation emitted as a color-corrected blackbody is compatible with the emitted spectrum.

The blue data point corresponds to the observation of the BH binary Cyg X-1 analyzed by \citet{Tomsick18}. We use the fit with an emissivity profile given in their table 3 and assume $M=20\msun$ \citep{Ziolkowski14} and $D=2.2$ kpc. Their value of the power-law normalization is an order of magnitude above the value seen on their fig.\ 6. Here we use the latter. We find $n_{\rm{cal}}/{\cal{R}} \lesssim 0.1 n_{\rm{fit}}$ with ${\cal R}\approx 0.2$ inferred from their fig.\ 6, which indicates some problems with the used assumptions and methods, similarly to the case of the hard state of GX 339--4.

We also include the results obtained for 17 Seyfert galaxies by \citet{Jiang19b}; see the black points in Figure 3. Here, $R_{\rm{in}}=R_{\rm{ISCO}}(a_*)$ is assumed, $a_*$ is fitted, the distances are based on the redshift for $H_0=73$\,km\,s$^{-1}$\,Mpc$^{-1}$, and {\tt{relxillD}} is used. We see a very large scatter, indicating problems with some of the assumptions and/or methods. However, based on their figures, we have found ${\cal R}<1$ for all of the fitted models, which indicates $n_{\rm{cal}}\ll n_{\rm{fit}}$ on average.

Summarizing, we find $n_{\rm{cal}}\ll n_{\rm{fit}}$ on average for the hard state and for Seyferts, and $n_{\rm{cal}}\gg n_{\rm{fit}}$ for the soft state of GX 339--4. This indicates some issues with the method and assumptions. We stress, however, that all $F_{\rm irr}$, $n$ and $\xi$ depend on the radius, and the fitted values of $\xi$ and $n$ should be considered as corresponding to some representative region of the accretion disk. Future models employing profiles of the fitted quantities are desirable.

\section{Discussion and Conclusions}

We have derived a powerful method to constrain the inner radii of accretion disks in accreting BH binaries based on the Stefan-Boltzmann law, $F_{\rm{irr}}={\sigma}T_{\rm{eff}}^4$. The constraint follows from considering the flux irradiating the innermost parts of the disk, which requires the presence of a quasi-thermal component at a color temperature higher than $T_{\rm{eff}}$ implied by the strength and broadening of the Fe K and reflection components. The apparent absence of this component with $kT_{\rm{col}}\sim 1$\,keV in the data accumulated so far is not compatible with the presence of a strongly irradiated disk close to the ISCO in high-luminosity hard states, e.g., in GX 339--4. This constraint has been overlooked before owing to the widespread use of reflection codes assuming low density and fitting the ionization parameter, $\xi\propto{F_{\rm{irr}}}/n$, which implicitly yields $F_{\rm{irr}}$ orders of magnitude below the fluxes typical for the luminous hard state, and consequently the effective temperatures of the irradiated media having values consistent with the observations even at $R_{\rm{in}}{\approx}R_{\rm{ISCO}}$.  

We have derived the above constraint by conservatively assuming that the disk is irradiated only by a small fraction of the primary flux, which is the case if there is a hot inner flow at $R<R_{\rm{in}}$. For pure coronal geometry, the irradiating flux at the disk surface is higher, making the constraint stronger. The constraint is also stronger in the presence of intrinsic disk dissipation, as well as in the geometry of a lamppost surrounded by a disk extending close to the ISCO, where the disk is irradiated by gravitationally focused primary radiation. On the other hand, our constraint would be satisfied if either the height of the lamppost were very high, e.g., several tens ${R_{\rm{g}}}$, or the corona had most of its dissipation at large radii, which would result in only weak irradiation of the inner disk but also in a relatively narrow Fe K line. Apart from those possibilities, our constraint implies the truncated disk geometry in luminous hard states.

Our results present one more case of a tension between a large number of spectral results showing very broad Fe K lines in the luminous hard state (see the references in Section \ref{intro}) and the results based on time lags (e.g., \citealt{DeMarco15}), modeling of type C quasi-periodic oscillations as precession of the inner hot disk (e.g., \citealt{Ingram16}), some spectral fits (e.g., \citealt{Basak16, Basak17}), energy balance \citep{PVZ18}, comprehensive modeling of both the spectral and timing features \citep{Mahmoud19}, and considering e$^\pm$ pair equilibrium and BH photon trapping in the lamppost case \citep{NZS16}. At present, we do not understand the origin of this tension. Our results also put in question the determinations of the BH spin done with models neglecting the effects of disk irradiation at high illuminating fluxes, possibly easing the tension with implications of the Laser Interferometer Gravitational-Wave Observatory (LIGO)/VIRGO measurements, e.g., \citet{Belczynski20}.

The Stefan-Boltzmann constraint becomes weaker at lower luminosities. But given the slow dependence of $T_{\rm{eff}}\propto{F_{\rm{irr}}}^{1/4}$, it is still quite powerful at $L$ of $\sim$1\% of the Eddington luminosity in BH binaries. The strict application of our formalism in the future would require an availability of a public reflection code valid at densities $>10^{19}$\,cm$^{-3}$. In the Seyfert case, the thermal feature from the irradiation occurs in the UV range. It is highly desirable to study the resulting constraints given the evidence for the disks extending close to the ISCO in those systems (e.g., \citealt{DeMarco13}).

In the second part of the Letter, we have derived a related method based on comparing the disk densities fitted (by other authors) using high-density reflection codes with those calculated by us from $n\propto\xi{F_{\rm{irr}}}$ using the fitted ionization parameter, the source luminosity and the disk inner radius. However, we have found significant discrepancies between the densities estimated using the two methods, preventing us from obtaining significant constraints on the disk inner radius as yet. In the hard state and for Seyferts, the fitted densities were higher than those estimated by us, and in the high-luminosity very high state of GX 339--4, where the assumed $R_{\rm{in}}\approx{R_{\rm{ISCO}}}$ is likely, the fitted density was much too low to be self-consistent. It is clear that more work is required to achieve a reasonable sensitivity of this method in constraining $R_{\rm{in}}$. Still, it is potentially very useful in estimating the parameters of accreting sources.

\section*{Acknowledgments}
We thank J. Garc{\'{\i}}a, J. Jiang, A. Nied{\'z}wiecki, M. Parker and M. Szanecki for valuable discussions, and the two referees for valuable suggestions. We benefited from discussions during Team Meetings of the International Space Science Institute (Bern), whose support we acknowledge. We also acknowledge support from the Polish National Science Centre under the grants 2015/18/A/ST9/00746 and 2015/17/B/ST9/03422, and the European Union's Horizon 2020 research and innovation programme under the Marie Sk{\l}odowska-Curie grant agreement No.\ 798726.

\bibliography{allbib}{}
\bibliographystyle{aasjournal}

\label{lastpage}
\end{document}